\documentclass[epsfig,12pt]{article}
\usepackage{epsfig}
\usepackage{graphicx}
\usepackage{array}
\usepackage{color}
\usepackage{bm}
\usepackage{cite}
\usepackage{geometry}
\usepackage{latexsym}
 \usepackage{amsmath, amssymb, amscd, xypic, graphicx}

\newcommand{\beq}{\begin{equation}}   
\newcommand{\eeq}{\end{equation}}
\newcommand{\beqn}{\begin{eqnarray}}   
\newcommand{\eeqn}{\end{eqnarray}}

\newcommand*\xbar[1]{%
 \kern0.5ex%
  \hbox{%
   \kern0.2ex%
      \vbox{%
      \hrule height 0.5pt 
      \kern0.5ex
      \hbox{%
        \kern-0.1em
        \ensuremath{#1}%
        \kern-0.1em
      }%
    }%
  }%
}

\newcommand{\gsim}{\lower.7ex\hbox{$
\;\stackrel{\textstyle>}{\sim}\;$}}
\newcommand{\lsim}{\lower.7ex\hbox{$
\;\stackrel{\textstyle<}{\sim}\;$}}
\begin{document}

\begin{titlepage}

\begin{flushright}
FTPI-MINN-14/15, UMN-TH-3338, NSF-KITP-14-060, IHES/P/14/20
\end{flushright}

\vspace{1cm}

\begin{center}
{  \Large \bf  Making supersymmetric connected  \\[2mm]
\boldmath{${\mathcal N}\!=\!(0,2)$}  sigma models 
}

\end{center}

\begin{center}
{\large
 Mikhail Shifman,$^{a,b,c}$ Arkady Vainshtein,$^{a,b,d}$ and Alexei Yung$^{\,b,e}$}
\end {center}

\vspace{1mm}

\begin{center}

$^{a}${\it  Department of Physics, University of Minnesota,
Minneapolis, MN 55455, USA}\\[1mm]
$^b${\it  William I. Fine Theoretical Physics Institute,
University of Minnesota,
Minneapolis, MN 55455, USA}\\[1mm]
$^{c}${\it Institut des Hautes \'{E}tudes Scientifiques, 35 Route de Chartres, 91440 Bures-sur-Yvette, France}\\[1mm]
$^{d}${\it Kavli Institute for Theoretical Physics, University of California,\\ Santa Barbara, CA 93106, USA}\\[1mm]
$^{e}${\it Petersburg Nuclear Physics Institute, Gatchina, Leningrad district,
188300, Russia}\\[1mm]

\end{center}

\vspace{0.6cm}

\begin{center}
{\large\bf Abstract}
\end{center}

We construct ``connected" (0,2) sigma models starting from $n$ copies of (2,2) CP$(N-1)$ models. 
General aspects of models of this type (known as $T+O$ deformations) had been previously studied in the context of
heterotic string theories. Our construction
presents a natural generalization of the nonminimally 
deformed (2,2) model with an extra (0,2) fermion superfield on tangent bundle 
${\rm T}\big[{\rm CP}(N\!-\!1)\!\times \!{\rm C^{1}}\big]$.\ We had thoroughly
analyzed the latter model previously, found the
exact $\beta$ function and
a spontaneous breaking of supersymmetry.
In contrast,  in certain connected sigma models the spontaneous breaking 
of supersymmetry disappears.
We study the connected sigma models in the large-$N$ limit finding supersymmetric vacua 
and determining the particle spectrum.
While the Witten index vanishes in all the models under consideration, in these special cases of
connected models one can use a permutation symmetry to define 
a modification  of  the Witten index which does not vanish. This
eliminates  the spontaneous breaking of supersymmetry. 
We then examine the exact $\beta$ functions of our connected (0,2) sigma models.

\end{titlepage}

\newpage

\tableofcontents

\newpage

\section{Introduction}

\begin{figure}[h]
\epsfxsize=4cm
\centerline{\epsfbox{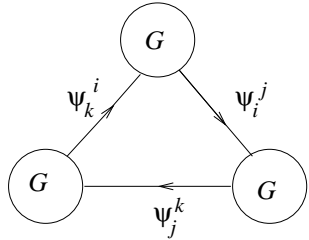}}
\caption{\small
SU$(N)\times$SU$(N)\times$SU$(N)$ Yang-Mills theory with bifundamental fermions.}
\label{f1}
\end{figure}

Quiver gauge theories in four dimensions are useful in various applications.
The most common in four dimensions are Yang-Mills theories of the type 
$$\mbox{
SU$(N_1)\times$SU$(N_2)\times$SU$(N_3)\!\times ...$}
$$ 
(the  ``nodes") with each factor group being cyclically connected to its neighbors
 by a set of bifundamental fermion fields transforming in the fundamental representation
of a given SU$(N)$ theory and in the antifundamental representation of its 
neighbor. These fermion fields can be represented in the quiver graph as arrows (see Fig.\,\ref{f1}).

Two-dimensional asymptotically free sigma models are long known to be excellent 
laboratories for modeling four-dimensional Yang-Mills theories.\footnote{\,It was 40 years ago that 
Polyakov emphasized \cite{poly} that asymptotically free two-dimensional sigma models
could present the best laboratory for the four-dimensional Yang-Mills theories. His prophecy came true. }  
The question we ask is whether one can construct an analog of quiver Yang-Mills in the context
of two-dimensional sigma models. Moreover, we require a part of supersymmetry 
to be preserved in this construction.

In answering the above question we can use, for guidance, 
previous work carried out in the context of heterotic string theories in which models
known as deformations  $T+O$  with $O$ being the
trivial bundle where discussed  \cite{A,B}. In many instances models
obtained are superconformal in the infrared (see e.g.  \cite{B}). 
Since we are interested in analogies with four-dimensional super-Yang-Mills
we would like to construct models with massive particle spectrum. 
To this end we turn our attention to a particular
case of the $T+O$ deformations starting from a ${\mathcal N}=(2,2)$ theory associated to
a product of $n$ two-dimensional  CP$(N\!\!-\!1$) models. A dynamical connection between them is realized
through one or several right-moving fermions from trivial tangent bundles.
Somewhat related constructions were discussed in recent publications \cite{Gadde:2013lxa,Gadde:2014ppa}.

Adding fermions in the bosonic sigma models (generally speaking, in a nonsupersymmetric manner) it is not difficult to ``connect" them. For, instance, one of the options
 is to couple two CP$(N\!-\!1)$ models, both nonsupersymmetric, as follows:
 \beq
{\mathcal L} =
 G \Big[\rule{0mm}{6mm}\partial^\mu \phi^{\dagger}\, \partial_\mu\phi
+i \bar\psi  \slash\!\!\!\nabla\psi\Big]
\!+\widetilde G \Big[\rule{0mm}{6mm}\partial^\mu \widetilde\phi^{\,\dagger}\, \partial_\mu\widetilde\phi
+i{\widetilde{\bar\psi}} \slash\!\!\!\nabla\widetilde\psi\Big] +
\beta\Big(\!R\,\psi_L^\dagger\psi_R\!\Big)\Big(\!\widetilde{R}\widetilde\psi_R^{\,\dagger}\widetilde\psi_L\!\Big)\,,
\label{minq}
\eeq
where the fields of the first CP$(N\!-\!1)$ are untilded, those of the second CP$(N\!-\!1)$ are tilded, $\beta$ is a coupling constant, 
$G$ and $R$ stand for the CP$(N-1)$ metric and Ricci tensor, respectively. Moreover,
$\nabla_\mu$ is the target-space covariant derivative.
The fermion fields $\psi_{L} $ and $\tilde\psi_R$ are chiral, left and right movers. 

As was mentioned, we would like
to find connected models with dynamical mass generation and a part of supersymmetry preserved.
Nonminimal \mbox{${\mathcal N}\! =\!(0,2)$} models seem to be an ideal starting point. 
 A nonminimal model to serve as our starting point
appeared as a low-energy theory on the world sheet of a non-Abelian BPS-saturated flux tube supported in an
${\mathcal N}=1$ four-dimensional Yang-Mills theory \cite{Edalati:2007vk,Shifman:2008wv}.
Our tool is the large-$N$ expansion, generalizing a number of results which had been obtained in the past in nonsupersymmetric and (2,2) supersymmetric $CP(N-1)$ models. 

One starts from the bulk four-dimensional theory with ${\mathcal N}\!=\!2$ supersymmetry which supports 1/2-BPS strings. Then
the low-energy theory on its world sheet has four supercharges and, thus, possesses ${\mathcal N}\!=\! (2,2)$
supersymmetry. The  target space  of the corresponding sigma model is CP$(N\!-\!1)$ \cite{nastrings}.
More exact, the full target space is  CP$(N\!-\!1)\!\times\!{\rm C}^1$ where ${\rm C}^1$ appears due to shifts of the string in transversal 
spatial directions. The bosonic fields living on ${\rm C}^1$ as well as their fermionic partners on the tangent bundle 
${\rm TC^{1}}$ are free fields, and  for this reason they usually are omitted from consideration.

If one slightly deforms the bulk theory, breaking ${\mathcal N}\!=\!2$ down to ${\mathcal N}\!=\!1$,
four supercharges in the bulk survive. For relatively small deformations BPS saturation remains valid and so does the target space of the two-dimensional sigma model. Now, the world-sheet model must have two, not four supercharges.
However, Zumino's theorem \cite{Zuminosigma} implies that given a K\"ahler target space any supersymmetric nonchiral model is automatically uplifted to ${\mathcal N}\!=\! (2,2)$, i.e. four supercharges.

Edalati and Tong \cite{Edalati:2007vk} noted that, in fact, the above deformation of the bulk theory gives rise to 
interaction for right-moving fermions living on ${\rm TC^{1}}$; in the bosonic background they start to mix with the
right-moving fermions on ${\rm TCP}(N\!-\!1)$. Thus, the bosonic target space stays the same CP$(N\!-\!1)$ (modulo free fields on ${\rm C}^{1}$)
while ${\mathcal N}\!=\! (2,2)$ is broken into ${\mathcal N}\!=\! (0,2)$ on the fermion tangent bundle 
${\rm T}\big[{\rm CP}(N\!-\!1)\times {\rm C^{1}}\big]$\,.
They also conjectured a certain ${\mathcal N}\!=\! (0,2)$ model on the string world sheet 
with the field content of ${\mathcal N}\!=\! (2,2)$ CP$(N\!-\!1)$ sigma model plus a $(0,2)$ spinor multiplet
defined on $C^1$. 
This nonminimal theory (in a slightly different form) was explicitly derived by Shifman and Yung 
\cite{Shifman:2008wv} 
from the analysis of the vortex solution. They also found a geometric formulation 
of this model, as well as its
large-$N$ solution \cite{L1}. This solution 
exhibits spontaneous breaking of supersymmetry, as it often happens in other ${\mathcal N}= (0,2)$ models
discussed in the literature. 

Heterotic two-dimensional models  [known as ${\mathcal N}\!\!= \!(0,2)$ supersymmetric sigma models] have two chiral supercharges, 
say, $Q_L$ and $Q_L^{\dagger}$, with the defining anticommutator
\beq
\{Q_L\,,\, Q_L^{\dagger}\} = 2 (H-P)\,.
\label{ai1}
\eeq
They were studied from the mathematical perspective \cite{west,Witten:2005px,bai2,bai3,Jia,bai4} as well as from the standpoint
of physical applications (see \cite{SYreview} and extensive references therein).

\begin{figure}[h]
\epsfxsize=4cm
\centerline{\epsfbox{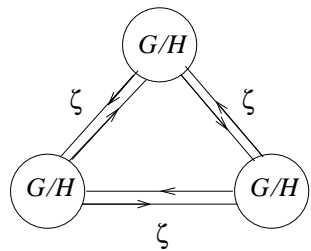}}
\caption{\small
T\big[CP$(N\!-\!1)\times\!$ CP$(N\!-\!1)\times$\! CP$(N\!-\!1)\!\times \!C^1\big]$ two-dimensional sigma model with one (0,2) fermion superfield. $G$ and $H$ are defined after Eq.\,(\ref{Bzeta}).}
\label{f2}
\end{figure}

The $(0,2)$ connected model we will construct has  the bosonic
target space 
\beq
{\rm CP}(N_1\!-\!1)\!\times\!{\rm CP}(N_2\!-\!1)\!\times ... \times\! {\rm CP}(N_n\!-\!1)\,. 
\eeq
As for the fermion fields they will live on the tangent bundles  of the type
\beq
{\rm T}\Big[{\rm CP}(N_1\!-\!1)\!\times\!{\rm CP}(N_2\!-\!1)\!\times ... \times\! {\rm CP}(N_n\!-\!1)\!\times\! C^1...\Big]\,.
\label{thfour}
\eeq
In the simplest version to be considered in Sec.~\ref{htm} 
there is a single connecting fermion $\zeta_R$ defined 
on the trivial tangent bundle ${\rm TC^1}$.
 All fields from 
CP$(N_p\!-\!1)$ interact with those from CP$(N_q\!-\!1)$ (for all $q,\,p = 1,2, ..., n$) through the coupling to the $(0,2)$ Fermi multiplet consisting of $\zeta_R$ and an auxiliary field.
The graph representation describing the case of ${\rm T}\big[{\rm CP}(N\!-\!1)\!\times \!{\rm CP}(N\!-\!1)\!\times \!{\rm CP}(N\!-\!1)\!\times\! C^1\big]$ is given in Fig.\,\ref{f2}.

\begin{figure}[h]
\epsfxsize=4cm
\centerline{\epsfbox{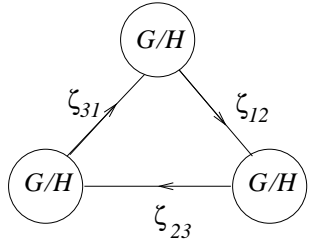}}
\caption{\small Graphic representation for the two-dimensional sigma model with the target space (\ref{6}).}
\label{f3}
\end{figure}

In Sec.\,\ref{otq} we construct a (0, 2) Lagrangian describing a connected sigma model with a cyclic graph of
the type given in Fig.\,\ref {f1}
with $n$ nodes and $n$ arrows  $(\zeta_R)_{j,j+1}$. Each arrow corresponds to its own (0, 2) fermion superfield (Fig.\,\ref{f3}), so that 
the target-space structure is as follows:
\beq
{\rm T}\Big[{\rm CP}(N_1\!-\!1)\!\times\! C^1\!\times \!{\rm CP}(N_2\!-\!1)\!\times \!C^1\!\times\! {\rm CP}(N_3\!-\!1)
\!\times\! C^1...\Big]\,.
\label{6}
\eeq

All models we consider have  ${\mathcal N}= (0, 2)$ supersymmetry at the Lagrangian level.  
In the leading order in $1/N$,
we select the models (choosing $N_{1},\,N_{2}, \ldots$ appropriately) where 
there is {\em no} spontaneous supersymmetry breaking. Once it happens in the leading $1/N$ order 
one can argue that restoration of supersymmetry is then an exact statement.

An important characteristic of the model associated with spontaneous breaking of supersymmetry is 
the  Witten index of the model \cite{Wind}. 
Spontaneous breaking can occur only
when this index vanishes. In Sec.\,\ref{WI} we show that 
Witten's index vanishes for all the models we consider.
However, one can introduce a modification of Witten index (a particular case of algebraic genera), following the same line 
of reasoning as in Sec.\,6 of  \cite{Witten:1982df}. In massless 
${\mathcal N}\!=\!1$  QED in four dimensions Tr\,$(-1)^F =0$; however,
Tr\,$C(-1)^F   = 4$, where $C$ stands for the $C$ parity. In our case, we can use flavor permutations to refine a modified index.

The paper is organized as follows. In Sec.\,\ref{tshn} we briefly review the variant of
the nonminimal heterotic modification of  ${\mathcal N}\!=\! (2,2)$  theory to be used in our study. In Secs.\,\ref{htm}
and \ref{otq} we fully specify the connected models with the target spaces (\ref{thfour}), solve them at large $N$ and determine the mass spectrum. Section \ref{WI} is devoted to the Witten index and its refinement. In Sec.\,\ref{befu} we obtain the exact $\beta$ function. Finally, Sec.\,\ref{concl}
summarizes our conclusions.

\section{The nonminimal heterotic modification\\ of \mbox{\boldmath{${\mathcal N}\!=\! (2,2)$}} theories: generalities}
\label{tshn}
 
The Lagrangian of generic ${\mathcal N}\!=\! (2,2)$ sigma model without torsion has the form
\beq
\mathcal{L} =\int\! d^{4}\theta \,K(\Phi, \Phi^{\dagger})\,,
\eeq
where the K\"ahler potential $K$ depends on the chiral superfields $\Phi^{i}$ and antichiral $\Phi^{\dagger \bar j}$.
Having in mind a manifold $M$ which is a direct product of $n$ manifolds $M_{F}$, $(F=1,\dots,n)$,
\beq
M=M_{1}\times M_{2}\times \dots \times M_{n}\,,
\eeq
we can rewrite ${\cal L}$ as a sum over manifolds,
\beq
\mathcal{L} =\sum_{F=1}^{n}\int d^{4}\theta \,K_{F}(\Phi_{F}, \Phi_{F}^{\dagger})\,.
\eeq
In terms of the (0, 2) superfields the field $\Phi^{i}$ decomposes as
\beq
\begin{split}
\Phi^{i}(x_{R}+2i\theta_{R}^{\dagger}\theta_{R},x_{L}-2i\theta_{L}^{\dagger}\theta_{L},\theta_{R},\theta_{L})\qquad \qquad\qquad\qquad ~~~\\[2mm]
=A^{i}(x_{R}+2i\theta_{R}^\dagger\theta_{R},x_{L}-2i\theta_{L}^{\dagger}\theta_{L},\theta_{R})+\sqrt{2}\,\theta_{L} B^{i}(x_{R}+2i\theta_{R}^\dagger\theta_{R},x_{L},\theta_{R}),
\end{split}
\label{decomp}
\eeq
where $x_{R,L}=x^{0}\pm x^{1}$ and the field $A^{i}$ represents the chiral supermultiplets which on the mass shell consists of  the  scalar 
field and left-moving fermion,
\beq
A^{i}=\phi^{i}(x_{R}+2i\theta^\dagger\theta,\,x_{L})+\sqrt{2}\,\theta\,\psi^{i}_L(x_{R}+2i\theta^\dagger\theta,\, x_{L})\,,
\label{Afield}
\eeq
while the field $B^{i}$ describes the Fermi supermultiplet which on mass shell contains only a right-moving fermion
($F^{i}$ is an auxiliary field),
\beq
B^{i}=\psi^{i}_R(x_{R}+2i\theta^\dagger\theta,\, x_{L})+\sqrt{2}\,\theta F^{i}_\psi(x_{R}+2i\theta^\dagger\theta,\, x_{L})\,.
\label{Bfield}
\eeq
Note a change in notation: in Eqs. (\ref{Afield}) and (\ref{Bfield}) $\theta_{R}$ is substituted by $\theta$ because 
we do not use $\theta_{L}$ in the (0,\,2) superspace.

The (0,\,2) heterotic modification is due to adding  one extra Fermi supermultiplet which we denote as ${\cal B}$,
\beq
{\mathcal B}=\zeta_R(x_{R}+2i\theta^\dagger\theta,\,x_{L})+\sqrt{2}\,\theta\,{\mathcal F}_\zeta(x_{R}+2i\theta^\dagger\theta,\, x_{L})\,,
\label{Bzeta}
\eeq
which interacts with the fields $A_{F}, B_{F}$ defined on each manifold $M_{F}$.
The introduction of this right-moving fermion does not change geometry of the original bosonic manifold.
Indeed, the manifold $M_{F}$ we deal with is a symmetric space associated with $G_{F}/H_{F}$ [in the case of CP($N\!-\!1$)
we have $G\!=\!{\rm SU}(N)$ and  $H\!=\!{\rm S}\left({\rm U}(N\!-\!1) \!\times \! {\rm U}(1)\right)$].
The additional field ${\cal B}$ is a singlet of the isometry group $G_{F}$, in contrast to $A_{F}^{i},\, B_{F}^{i}$.
Therefore, its interaction does not modify the isometry group.

The  Lagrangian of the model is
\beqn
\mathcal{L}
\!&=&\! \sum_{F=1}^{n}\left\{ -\frac{1}{2} \,{\rm Re} \!\int \!  d \theta\,  G^{F}_{i\bar j}(A_{F}, A_{F}^\dagger) (\xbar{\!D}A_{F}^{\dagger \bar j})\left(i\partial_{R} A_{F}^i -2\kappa_{F} \,\mathcal{B} B_{F}^i\right)
\right .
\nonumber\\[1mm]
&~&~~~~~+\,
\left . \frac{1}{2}\!\int \!d^2 \theta
\,Z_{F}\,G^{F}_{i\bar j}(A_{F}, A_{F}^\dagger)\, B_{F}^{\dagger\bar j} B_{F}^i \right\} + \frac{1}{2}\!\int \!d^2 \theta\, {\mathcal Z} \,\mathcal{B}^\dagger \mathcal{B}\,,
\label{sfLa}
\eeqn
where $G^{F}_{i\bar j}=\partial^{2} K^{F}/\partial A^{i}\partial A^{\dagger\,\bar j}$  is the K\"ahler metric of $M_{F}$, $\partial_{L,R} =\partial_{x^{0}}\pm \partial_{x^{1}}$ and $D = \partial_{\theta} -i \bar\theta \partial_L$, $\xbar D = -\partial_{\bar\theta} +i \theta \partial_L$.

Moreover, the parameters  $\kappa_{F}$ are the deformation parameters, and $Z_{F},\,{\cal Z}$ are wave function renormalization 
factors for $B_{F}^{i},\,{\cal B}$ fields. When all $\kappa_{F}=0$, the field ${\cal B}$ becomes a sterile field, and the (2,\,2) supersymmetry in the nontrivial sector
is restored. (The $Z_{F}$ factors do not run at $\kappa=0$ and can be taken to be 1.)

In components we have
\beqn
&&\hspace{-12mm}\mathcal{L} =\sum_{F=1}^{n}\left\{
G^{F}_{i\bar j}\left[\partial_{L}\phi_{F}^{\dagger \bar j} \partial_{R}\phi_{F}^{i}+\psi_{FL}^{\dagger \bar j}\,i\nabla_{\!R}\,\psi_{FL}^{i}
+ Z_{F}\,\psi_{FR}^{\dagger \bar j}\,i\nabla_{\!L}\psi_{FR}^{i}\right]
 \right . \nonumber\\[1mm]
&&\left . ~~+ Z_{F}R^{F}_{i{\bar j} k {\bar l}}\,\psi_{FL}^{\dagger \bar j}\psi_{FL}^{i} \,\psi_{FR}^{\dagger \bar l}\psi_{FR}^{k} +\!
\left[\kappa_{F}\, \zeta_R  \,G^{F}_{i\bar j}\big( i\,\partial_{L}\phi_{F}^{\dagger \bar j}\big)\psi_{FR}^{i}
+{\rm H.c.}\right] \! \right .\label{components}\\[1mm]
&&\left . ~~+\frac{|\kappa_{F} |^2}{Z_{F}} \zeta_R^\dagger\, \zeta_R
\big(G^{F}_{i\bar j}\,  \psi_{FL}^{\dagger \bar j}\psi_{FL}^{i}\big)\right\}-\,\frac{1}{\mathcal Z}\left|
\sum_{F=1}^{n}\kappa_{F}G^{F}_{i\bar j}\psi_{FL}^{\dagger \bar j}\psi_{FR}^{i} \right |^{2}
+{\mathcal Z}\,\zeta_R^\dagger \, i\partial_L \, \zeta_R \,.\nonumber
\eeqn
Here $\nabla_{\!L,R}$ are covariant derivatives, 
\beq
\nabla_{\!L,R}\,\psi_{R,L}^{i}=\partial_{L,R}\,\psi_{R,L}^{i}+\Gamma^{i}_{kl}\,\partial_{L,R}\,\phi^{k}\,\psi_{R,L}^{l}\,. 
\label{ai9}
\eeq

\section{The simplest connection of \boldmath{$n$} CP\boldmath{$(N\!-\!1)$} models}
\label{htm}

The general idea is to choose the manifold $M_{F}={\rm CP}(N\!-\!1)$ for all $F=1,\dots,n$  and couple all sectors through the field(s)
${\mathcal B}$ (see Fig.\,\ref{f2}). This coupling can be realized in various forms. The simplest one is a universal coupling of a single ${\mathcal B}$ field to all CP$(N\!-\!1)$ sectors with one and the same coupling constant $\kappa_{F}=\kappa,~F=1,\dots,n$. The right-moving fermions then
live on the tangent bundle of the form 
\beq
{\rm T}\Big[\{{\rm CP}(N-1)\}^{n}\times C^1\Big]\,.
\label{ts1}
\eeq 
Let us first discuss this version and then
move on to consider more elaborate models with the same underlying idea and a number of different 
${\mathcal B}$ superfields (Fig.\,\ref{f3}). 

The geometric formulation of the models is given then by Eqs.\,(\ref{sfLa}) and (\ref{components}) where the metric 
$G^{F}_{i\bar j}$\,,
 the heterotic coupling $\kappa_{F}$, and the wave function factors 
 $Z_{F}$ are the same for each $M_{F}={\rm CP}(N\!-\!1)$. The field
indices $i,{\bar j}$
run from 1 to $N\!-\!1$, and the explicit expressions for the CP($N\!-\!1$) metric and related objects are of the form,
\beqn
&& K=\frac{2}{g^{2}}\,\log \chi\,,\qquad \qquad \qquad \qquad ~~~~~~\chi=1+\sum_{m}^{N-1}\phi^{\dagger\, m}\phi^{m}\,
 \label{212},\\
&& G_{i\bar j}=\frac{2}{g^{2}}\Bigg(\frac{\delta_{i\bar j}}{\chi}-\frac{\phi^{\dagger\,i}\phi^{\bar j}}{\chi^{2}}\Bigg)\,,\qquad
\qquad
~G^{i\bar j}=\frac{g^{2}}{2}\,\chi\,\Big(\delta^{i\bar j}+\phi^{i}\phi^{\dagger \,\bar j}\Big)\,, \nonumber\\[1mm]
&& \Gamma^{i}_{kl}=-\frac{\delta^{i}_{k}\,\phi^{\dagger\,l}+\delta^{i}_{l}\,\phi^{\dagger\,k}}{\chi}\,,
\qquad \qquad\quad ~\Gamma^{\bar i}_{\bar k \bar l}=-\frac{\delta^{\bar i}_{\bar k}\,\phi^{ \bar l}+\delta^{\bar i}_{\bar l}\,\phi^{ \bar l}}{\chi}\,,\nonumber \\[1mm]
&& R_{i{\bar j}k{\bar l}}=-\frac{g^{2}}{2}\Big(G_{i\bar j}G_{k\bar l}+G_{k\bar j}G_{i\bar l}\Big)\,,\quad
 ~R_{i\bar j}= - G^{k \bar j}R_{i{\bar j}k{\bar l}}=\frac{g^{2}N}{2}\, G_{i\bar j}\,.\nonumber
\eeqn

The analogs of the gauge couplings $1/g^2$ are hidden in the metric tensors $G_{i\bar j}$, see Eq.\,(\ref{212}).
These couplings can be complexified by including $\theta$ terms,
\beq 
\frac{1}{g^{2}}\Longrightarrow \frac{1}{g^{2}}+ i\,\frac{\theta}{4\pi}\,.
\eeq
Later we will use such complexification to our benefit. 

The symmetry of the model is $\big[{\rm SU}(N)\big]^{n}$.
With our choice of all parameters,  the model
 acquires an additional flavor
$Z_n$ symmetry corresponding to interchanging different-$F$ fields.\footnote{\,This $Z_n$, which has no continuous analog, is not to be confused with the axial $Z_N$ for each flavor which is a remnant of the continuous classical $R$ symmetry}
More exactly, we define the  flavor symmetry as follows. Assume that the real parts
of $1/g^2$  are the same for all $n$ $CP(N-1)$ factors, while the $\theta$ terms take the values 
$0, 2\pi, 4\pi, ..., 2\pi (n-1)$.  Since for all $\theta= 2\pi \times$integer, physics is the same,
the permutation symmetry will  be valid with the appropriate choice of vacua.

\subsection {Gauged formulation of modified \boldmath{$\big[{\rm CP}(N\!-\!1)\big]^{n}\! \!\times\! C^{1}$}}

The gauged formulation is defined by the groups $G$ and $H$ entering into the $G/H$ 
symmetric space under consideration. In the CP$(N\!-\!1)$ case $G={\rm SU}(N)$ and
the gauged formulation has the form \cite{Edalati:2007vk},
\beqn
{\mathcal L} 
&=&  \sum_{F=1}^{n}\Bigg\{
{\mathcal D}_\mu n^{\dagger}_{Fi} {\mathcal D}_\mu n_{F}^i - \big(2|\sigma_{F}|^2+D_{F}\big) n^{\dagger}_{Fi}n_{F}^i
+\xi_{FLj}^\dagger\,i{\mathcal D}_{\!R} \,\xi^j_{FL}+Z\,\xi_{FRj}^\dagger\,i{\mathcal D}_{\!L} \,\xi^j_{FR} \nonumber \\[2mm]
&~&+
\left[\sqrt{2Z}\, i\,\sigma_{F} \xi_{FRj}^\dagger \xi^j_{FL} +
 \sqrt{2}\,i\,n^{\dagger}_{Fj} \lambda_{FR}\xi_{FL}^j + \sqrt{2Z}\,i\,n^{\dagger}_{Fj} \lambda_{FL}\xi_{FR}^j +{\rm H.c.}\right]
\nonumber\\[2mm]
&&+\frac{2}{g^{2}}\,D_{F}-\frac{\theta}{2\pi}\,\epsilon^{\mu\nu}\partial_{\mu}A_{F\nu}
 + \left[ \frac{2\kappa}{g^{2}}\sqrt{2{\cal Z}}\,i\,\zeta_R\,{\lambda}^{\dagger}_{FL}\,
  + {\rm H.c.}\right]\Bigg\} - 8\,\frac{|\kappa|^{2}}{g^{4}}\Big | \sum_{F=1}^{n} \sigma_{F}\Big |^2
 \nonumber\\[1mm]
 &~&+\,{\cal Z}\,\zeta^{\dagger}_R \, i
\partial_L\, \zeta_R\, \,.
\label{13}
\eeqn
Here $n_{F}^i$ $(i=1,\dots,N)$ is the complex scalar field in the fundamental representation of SU$(N)$, $\xi^i_{FL},~\xi^{i}_{FR}$
are its fermion superpartners in unbroken ${\cal N}\!=\!(2,\,2)$. 
The covariant derivatives, defined as ${\mathcal D}_\mu n_{F}^i =(\partial_{\mu}-iA_{F\mu})n_{F}^i$, 
contain auxiliary Abelian  gauge fields $A_{F\mu}$.
 The gauge field (2,\,2) superpartners $D_{F}$, $\sigma_{F}$, and $\lambda_{FL}, \lambda_{FR}$ are other auxiliary fields  which implement the constraints 
\beq 
 n^\dagger_{Fi} n_{F}^{i}=\frac{2}{g^{2}}\,, \qquad n_{Fj}^\dagger \xi^{j}_{FL} =0\,, \qquad \sqrt{Z}n^\dagger_{Fj} \xi^{j}_{FR} =
2\sqrt{{\cal Z}}\,\frac{\kappa^{*}}{g^{2}}\,\zeta^{\dagger}_R\,.
\eeq

\subsection{Large\,-\boldmath{$N$} solution}

It is easy to solve the theory (\ref{13}) in the 't Hooft limit, using the method of  \cite{gfw,W93}. In fact, at $N\to \infty$, only one-loop diagrams survive, as explained in detail in \cite{gfw,W93} (and in \cite{L1} in application to the heterotic model (\ref{13}) under consideration). 

The running  of the wave function factors $Z$ and ${\cal Z}$ shows up only in the $1/N$ corrections 
(see Sec.\,\ref{befu} and \cite{CCSV} for further details of running), so 
in the leading \mbox{large-$N$} approximation we put $Z\!=\!{\cal Z}\!=\!1$.
 Note also that in each CP($N\!-\!1$) sector, 
the auxiliary fields $A_{\mu}$, $D$, $\sigma$, and $\lambda_{L,R}$ form a supermultiplet of ${\cal N}\!=\!(2,2)$. 
The heterotic modification decomposes it into two (0,\,2) multiplets: a vector one, containing $A_{R}$, $\lambda_{R}$,  $\lambda^{\dagger}_{R}$, $D$,
and a chiral multiplet with $\sigma$ and $\lambda^{\dagger}_{L}$ fields.

To determine the vacuum structure it is sufficient to set $A_\mu =0$ and $\lambda_{L,R}=0$,
and treat $D$ and $\sigma$ as constant background fields, the critical values of which determine
the vacuum energy density. The Lagrangian (\ref{13}) is quadratic in both, the $n$ fields and their fermion superpartners $\xi$. Therefore, they can be integrated out exactly. This yields
\beq
\prod_{F=1}^{n}
\frac{{\rm Det}\left(\rule{0mm}{4mm} -\partial_\alpha^2 - 2 |\sigma_{F}|^2
\right)^N}{{\rm Det}\left(\rule{0mm}{4mm}-\partial_\alpha^2 - D_{F}- 2 |\sigma_{F}|^2
\right)^N}\,\,.
\label{detdet}
\eeq
 The denominator comes from the boson loop while the numerator from the fermion loop.
 Although $\sigma$ is generically complex its phase can always  be absorbed  in the $\theta$ term in 
 Eq.\,(\ref{13}) by U(1) rotation of fermion fields. 
 The one-loop graph contributions in (\ref{detdet}) are simply calculable,
 \beq
V_{\rm one\,loop}\! =\frac{N}{4\pi}\sum_{F=1}^{n}\left[(D_{F}+2|\sigma_{F}|^2
)\Big(\log \frac{M_{\rm uv}^{2}}{D_{F}+2|\sigma_{F}|^2} +1\Big) - 2|\sigma_{F}|^2\Big(\log\frac{M_{\rm uv}^{2}}{2|\sigma_{F}|^2}+1\Big)
\right],
\label{veffp}
\eeq
where $M_{\rm uv}$ is the ultraviolet cutoff.
Then, the effective potential takes the form
\beq
\begin{split}
V_{\rm eff} &= V_{\rm one\,loop}-\frac{2}{g^{2}}\sum_{F=1}^{n}D_{F} +8\,\frac{|\kappa|^{2}|}{g^{4}}\,\Big |\sum_{F=1}^{n}\sigma_{F} \Big |^2 \\[1mm]
&=\frac{N}{4\pi}\Bigg\{\sum_{F=1}^{n}\left[D_{F}\Big(\log \frac{\Lambda^{2}}{D_{F}+2|\sigma_{F}|^2} +1\Big) 
+ 2|\sigma_{F}|^2\log\frac{2|\sigma_{F}|^2}{D_{F}+2|\sigma_{F}|^2}
\right]\\[1mm]
&~~~+2u\,\Big |\sum_{F=1}^{n}\sigma_{F} \Big |^2\Bigg\}
\,.
\end{split}
\label{18}
\eeq
Here we introduced the scaling parameter $\Lambda$,
\beq
\Lambda=M_{\rm uv}e^{-4\pi/Ng^{2}}
\label{lambda}
\eeq
and the heterotic deformation parameter $u$,
\beq
u=\frac{16\pi |\kappa|^{2}}{Ng^{4}}\,.
\eeq

The auxiliary field $D_{F}$ can be excluded by the condition $\partial V_{\rm eff}/\partial D_{F}=0$\,,
\beq
D_{F}=\Lambda^{2}-2|\sigma_{F}|^{2}\,,
\label{DF}
\eeq
and the effective potential for the $\sigma_{F}$ fields becomes
\beq
V_{\rm eff} (\sigma)=\frac{N}{4\pi}\,\Bigg\{\sum_{F=1}^{n}\left[\Lambda^{2}-2|\sigma_{F}|^{2}
+ 2|\sigma_{F}|^2\log\frac{2|\sigma_{F}|^2}{\Lambda^2}
\right]+2u\,\Big |\sum_{F=1}^{n}\sigma_{F} \Big |^2\Bigg\}\,.
\label{Veff}
\eeq

\subsection{Vacuum structure}
\label{vst}

Let us start with undeformed case when the heterotic parameter $u=0$.
Then we have just $n$ disconnected copies
of the (2,\,2) CP$(N\!-\!1)$ sigma models. Each of these copies has $N$  supersymmetric vacua \cite{gfw,W93},
\beq
\langle D_F\rangle =0,\quad \langle \sigma_{F}\rangle_{k_{F}} = \frac{\Lambda}{\sqrt 2}\exp\Big(\frac{2\pi \,i\, \ell_F}{N}\Big), \quad \ell_{F}=0,1, ..., (N\!-\!1)\,.
\label{nou}
\eeq

The value of $|\sigma_{F}|$ follows from minimization of $V_{\rm eff}$ in Eq.\,(\ref{Veff}), the value of $D_{F}$ is then given by Eq.\,(\ref{DF}).
How did the phase factor of $\sigma_{F}$ appear in Eq.\,(\ref{nou})\,? The phase of the vacuum value of $\sigma$ can always be absorbed in the $\theta$ term which, in turn, can be hidden in the definition (\ref{lambda}) of $\Lambda$ (we chose $\theta\!=\!0$ for simplicity). Given the physical $2\pi$ periodicity in $\theta$, one arrives at the expression for $\sigma_{F}$ presented in (\ref{nou}). The multivaluedness of the vacuum expectation value of 
$\sigma$ is the same as that in the condensate $\langle \xi^\dagger_L \xi_R\rangle$.

The next step to consider is the case of one CP$(N\!-\!1)$, $n=1$, with nonvanishing heterotic parameter $u$. 
This was done in detail in Ref.\cite{L1}. The critical values are
\beqn
&&\langle \sigma\rangle_{k} = \frac{\Lambda}{\sqrt 2}\exp\left(-\frac{u}{2}+\frac{2\pi \,i\, \ell}{N}\right),\quad \ell= 0,1, ..., (N-1)\,,
\nonumber\\[2mm]
&& \langle D\rangle =  \Lambda^2 \left(1\!-\!\,e^{-u}\right),\quad 
\langle V_{\rm eff}\rangle=\frac{N}{4\pi}\,\langle D\rangle= \frac{N}{4\pi}\,\Lambda^2 \left(1\!-\!\,e^{-u}\right)\,.
\label{20}
\eeqn
The fact that the vacuum energy density $\langle V_{\rm eff}\rangle \neq 0$ for $u\neq 0$ indicates that (0,\,2) supersymmetry is spontaneously broken. Of course, this implies the emergence of a massless Goldstino, its
determination can be found in  \cite{L1}.

Now, let us turn to the quiverlike theories with   $n\!>\!1$ and show that for $n\!>\!1$, supersymmetric vacua appear.

Unbroken supersymmetry implies that in the vacuum, $\langle D_{F}\rangle =0$ for all $F$; then Eq.\,(\ref{DF}) fixes
$|\langle \sigma_{F}\rangle|=\Lambda/\sqrt{2}$ also for all $F$. Thus, in the supersymmetric vacua, $\langle \sigma_{F} \rangle$ 
are the same as in the undeformed case and given by Eq.\,(\ref{nou}).
The vacuum energy density (\ref{Veff}) 
at this value of $|\langle \sigma_{F}\rangle |$ is given by $u\,|\sum \sigma_{F}|^{2}$ and
vanishes  when
\beq
\sum_{F=1}^n \exp\left(\frac{2\pi \,i\, \ell_F}{N}\right) =0 \,.
\label{cond}
\eeq
Take, for example, $n=2$ where the total number of ``prevacua" is $N^{2}$. The condition (\ref{cond}) is satisfied if $|\ell_{2}-\ell_{1}|=N/2$. Of course, this is possible only for even $N$. We see the occurrence of $N$ supersymmetric vacua, $\ell_{1}=0,\dots, (N-1)$. In the remaining $N(N-1)$ would-be vacua supersymmetry is spontaneously broken. These vacua have nonvanishing energy and are cosmologically unstable. In Sec.\,\ref{WI} we argue
that the existence of supersymmetric vacua in our models extends beyond the leading $1/N$ approximation.
This is an exact statement. 

The absence of the spontaneous supersymmetry  breaking in the heterotic (0,\,2) theories is not a new phenomenon.
The (0,\,2) theories with supersymmetric vacua were recently discussed in Refs.\cite{Gadde:2013lxa,Gadde:2014ppa}. Classes of theories
such as  the (0,\,2) Landau-Ginzburg models, as well as (0,\,2) gauged linear sigma model (GLSM)
constructions of heterotic string vacua --- with supersymmetric vacua and superconformal regime in the IR --- had been also considered in the past. 
Their dynamics is quite different from what
we observe in our models. 

The resurgence of the supersymmetric vacua due to (\ref{cond}) can be understood from different angles. 
To this end, let us start discussing the mass spectrum of the models.

\subsection{Mass spectrum}
\subsubsection{Undeformed theory}
Let us start, again, with the undeformed case when the heterotic parameter \mbox{$u=0$}.
Then the right-moving fermion $\zeta_{R}$ represents a sterile massless field, and we have \mbox{${\cal N}\!=\!(2,2)$}
supersymmetry in the each $M_{F}={\rm CP}(N\!-\!1)$ sector. The supersymmetry is unbroken and the mass spectrum
at large $N$ is well known \cite{gfw,W93}.  The fundamentals of SU($N$), i.e., the fields $n^{i},~\xi^{i}$, get masses 
\beq
m_{n}=m_{\xi}=\sqrt{2}\,|\langle \sigma \rangle |=\Lambda\,,
\eeq
as it is visible from Eq.\,(\ref{13}). This means that strong interaction in the infrared produces extra states as compared 
to the original Lagrangian of the sigma model. This leads to the linear representation of SU($N$) and nonvanishing masses.

In addition,
the kinetic terms for the gauge $A_\mu$ field and its (2,\,2) superpartners $\sigma$ and $\lambda$ as well as the Yukawa 
$\sigma \lambda \lambda$ coupling are dynamically generated
at one loop in much the same way as
in \cite{gfw,W93},
\beqn
{\cal L}^{\rm kin}_{\rm one\,loop}&&\!\!\!\!\!\!\!\!\!=\!\frac{N}{4\pi \rho^{2}}\Big[\! -\frac 1 4 F_{\mu\nu}F^{\mu\nu}\!+\frac 1 2 \,\partial_{\mu }\rho\,\partial^{\mu}\rho
+\!\lambda_{L}^{\dagger}i\partial_{R}\lambda_{L}+\!\lambda_{R}^{\dagger}i\partial_{L}\lambda_{R} \nonumber\\[2mm]
&&\!\!\!+2i\rho \big(e^{i\alpha}\lambda_{R}^{\dagger}\lambda_{L}
\!-\!e^{-i\alpha}\lambda_{L}^{\dagger}\lambda_{R}\big)\Big]+\frac{N}{4\pi}\Big[\,\frac 1 2 \,\partial_{\mu }\alpha\,\partial^{\mu}\alpha +2\alpha \epsilon^{\mu\nu}\partial_{\mu}A_{\nu}\Big]\,.
\label{K1}
\eeqn
Here we represent the complex field $\sigma$ in terms of the real fields $\rho$ and $\alpha$,
\beq
\sigma=\frac{1}{\sqrt{2}}\,\rho \, e^{i\alpha}\,,
\eeq
i.e., the modulus and the phase of $\sigma$. 

The Yukawa $\sigma \lambda \lambda$ coupling generates mass for the $\lambda$ field,
$
m_{\lambda}=2\langle \rho \rangle =2\Lambda\,.
$
To read off this mass one should substitute fields $\rho$ and $\alpha$ in Eq.\,(\ref{K1}) by their vacuum values,
\beq
\langle \rho \rangle =\Lambda\,, \qquad \langle \alpha \rangle =\frac{2\pi}{N}\,\ell\,,
\eeq
and introduce canonically normalized fields
\beq
\tilde \lambda_{L}=\sqrt{\frac{4\pi}{N}}\, \frac{e^{i\langle\alpha\rangle/2}}{ \langle \rho \rangle}
\,\lambda_{L}\,, \qquad
\tilde \lambda_{R}=\sqrt{\frac{4\pi}{N}}\, \frac{e^{-i\langle\alpha\rangle/2}}{ \langle \rho \rangle} 
\,\lambda_{R}\,.
\label{tildelambda}
\eeq
The same mass $m_{\rho}=m_{\lambda}$ follows for the $\rho$ field from expansion of $V_{\rm eff}$ in $(\rho\! -\!\langle \rho \rangle)$ [see 
Eq.\,(\ref{Veff})] at $u=0$.

To show  that the gauge field has the same mass note that
equations of motion relate the deviation $\alpha\!-\!\langle \alpha \rangle$ to the gauge field, 
\beq
\alpha-\langle \alpha \rangle =\frac{1}{2\rho^{2}}\,\epsilon^{\mu\nu}\partial_{\mu}A_{\nu}\,,
\eeq
and lead to
\beq
m_{\rm ph}=m_{\rho}=m_{\lambda}=2\langle \rho \rangle =2\,\Lambda\,.
\eeq
A crucial feature of the model is that  the photon field $A_\mu$, in addition to the kinetic term, acquires a nonvanishing 
mass: the presence of massless fermion fields in the model shifts the pole in the photon propagator away from zero. 
 Thus, the gauge (2,\,2) multiplet $A_{\mu}, \rho,\lambda$ becomes propagating 
with the mass $m_{\rm ph}$.

Consequences of a massless vs.\ massive photon in two dimensions are radically different. A massless gauge field in two dimensions [bosonic CP$(N\!-\!1)$] implie confinement of charged particles, while the massive one [supersymmetric CP$(N\!-\!1)$] does not confine \cite{gfw,W93}. In one-to-one correspondence with this 
is the existence of $N$ degenerate vacua in the nonconfining case.
In the confining case [i.e., massless gauge field, as in the bosonic CP$(N\!-\!1)$] one of these vacua remains genuine while the remaining $N\!-\!1$ are uplifted and become quasistable states \cite{GSY05}.
\subsubsection{Deformed \boldmath{${\rm CP}(N\!-\!1)$}}

Now let us consider one  ${\rm CP}(N\!-\!1)$, $n=1$, with nonvanishing heterotic parameter $u$.
In this case supersymmetry is broken, but all $N$ vacua (see the second line in (\ref{20})) remain degenerate.
This degeneracy reflects spontaneous breaking of a $Z_N$ symmetry present in the model. Hence, the $n^{i}$ particles should remain unconfined. Correspondingly, the photon becomes massive much in the same way as in the 
(2,2) model. 

The masses of the fundamental fields $n$ and $\xi$ are different,
\beq
m_n =\Lambda ,\quad m_\xi = \Lambda e^{-u/2}
\eeq
Other particle masses change with $u$, see \cite{L1} for a detailed derivation.
At small $u$ they are close  to $2\Lambda$, with the massless Goldstino which predominantly coincides with $\zeta_{R}$
with a small admixture of $\lambda_{R}$\,. At large $u$ the Goldstino becomes predominantly $\lambda_{R}$ while
$\zeta_{R}$ together with $\lambda_{L}$ constitute two massive states with a large mass
\beq
 m_{\lambda_{L}, \zeta_{R}}=\Lambda\sqrt{u}\,.
 \eeq
The gauge and $\rho$ fields become light with masses
\beq
m_{\rm ph}=\sqrt{6}\Lambda e^{-u/2}\,,\qquad m_{\rho}=2\sqrt{3}\Lambda e^{-u/2}\,.
\eeq

\subsubsection{Mass spectrum at \boldmath{$u\neq 0$} }

Let us find  the mass spectrum of the quiverlike theory in more detail. For simplicity we consider the case $n=2$ and even $N$.
The vacuum values of the fields in the supersymmetric vacua in this case are
\beq
\langle \rho_{1}\rangle=\langle \rho_{2}\rangle=\Lambda\,,\quad \langle \alpha_{1}\rangle=\frac{2\pi}{N}\,k\,,\quad 
\langle\alpha_{2}\rangle=\langle \alpha_{1}\rangle +\pi\,.
\eeq
The masses of the bosonic and fermionic fields $n^i_F$ and $\xi^i_F$   
determined by the above vacuum expectation values are
\beq
m_{n}^{F}=m_{\xi}^{F}=\langle \rho_{F}\rangle=\Lambda\,,\quad (F=1,2) \,.
\label{nmass22}
\eeq
Note, that they are the same for all ``flavors'' $F$, do not depend on the deformation parameter $u$,
 and  bosons and fermions remain degenerate. Note also that the $D$-term constraints are effectively lifted.
For example, we have  $2nN$ real degrees of freedom in the bosonic sector rather than $2n\,(N\!-\!1)$ seen
quasiclassically.

Let us consider now the effective Lagrangian for fields of two gauge multiplets and the right-moving
fermion $\zeta_{R}$.
Combining Eqs.\,(\ref{Veff}) and (\ref{K1}) as well as the fermionic part of the heterotic deformation 
from Eq.\,(\ref{13}) we have
\beqn
{\cal L}_{\rm eff}&&\!\!\!\!\!\!\!\!\!=\zeta_{R}^{\dagger}i\partial_{L}\zeta_{R}+\frac{N}{4\pi}\!\sum_{F=1,2}\!\Bigg\{\frac{1}{ \rho_{F}^{2}}\Big[\! -\frac 1 4 F_{F\mu\nu}F_{F}^{\mu\nu}\!\!+\frac 1 2 \,\partial_{\mu }\rho_{F}\partial^{\mu}\!\rho_{F}
\!+\!\lambda_{FL}^{\dagger}i\partial_{R}\lambda_{FL}\!+\!\lambda_{FR}^{\dagger}i\partial_{L}\lambda_{FR}\nonumber\\[2mm]
&&\!\!+\,2i\rho_{F}\big(e^{i\alpha_{F}}\lambda_{FR}^{\dagger}\lambda_{FL}\!
-e^{-i\alpha_{F}}\lambda_{FL}^{\dagger}\lambda_{FR}\big)\Big]+\Big[\,\frac 1 2 \,\partial_{\mu }\alpha_{F}\,\partial^{\mu}\alpha_{F} +2\alpha_{F} \epsilon^{\mu\nu}\partial_{\mu}A_{F\nu}\Big] \label{Leff}\\[2mm]
&&\!\!\!-\left[\Lambda^{2}-\rho_{F}^{2}
+ \rho_{F}^2\log\frac{\rho_{F}^2}{\Lambda^2}
\right]+\sqrt{\frac{8\pi u}{N}}\left[ i\,\zeta_R\,{\lambda}^{\dagger}_{FL}\,
  + {\rm H.c.}\right]\Bigg\}
-u\,\frac{N}{4\pi}\Big |\!\sum_{F=1,2}\rho_{F}e^{i\alpha_{F}} \Big |^2\,.\nonumber
\eeqn
Here we assumed that the parameter $\kappa$ is real, its phase can be absorbed into a field redefinition of $\zeta_{R}$.

Let us emphasize that this ${\cal L}_{\rm eff}$ is exact for constructing the large-$N$ expansion for terms up to the second order in derivative. For fermions each fermionic field should be counted as a square root of derivative. The only term missing in Eq.\,(\ref{Leff}) is of the fourth order in 
$\lambda$\,. It does not contribute in our leading-$N$ calculation but will enter for $1/N$ corrections. (These terms can be found in \cite{bsy4}.)

Now let us consider the effect of the heterotic modification for the mass spectrum.
Consider first the bosonic masses. Expanding the modification term [the last one in Eq.\,(\ref{Leff})] near
the vacuum values we get
\beq
\begin{split}
&-u\,\frac{N}{4\pi}\Big |\!\sum_{F=1,2}\rho_{F}\,e^{i\alpha_{F}} \Big |^2\!=-4u\,\frac{N}{4\pi}\left[\frac 1 2 \,\Big(\frac{\tilde \rho_{1}-\tilde \rho_{2}}{\sqrt{2}}\Big)^{2}
+\frac 1 2\,\Big(\frac{\tilde \alpha_{1}-\tilde \alpha_{2}}{\sqrt{2}}\Big)^{2}\right]\,,\\[1mm]
&\quad \tilde\rho_{F}=\rho_{F}-\langle \rho_{F}\rangle\,,\quad 
\tilde\alpha_{F}=\alpha_{F}-\langle \alpha_{F}\rangle\,.
\end{split}
\eeq
This means that the mass of the $(\tilde \rho_{1}+\tilde \rho_{2})/\sqrt{2}$ field as well as the mass of $(\tilde \alpha_{1}+\tilde \alpha_{2})/\sqrt{2}$
and the corresponding gauge field combination is not modified by the heterotic coupling,
\beq
m\big[(\tilde \rho_{1}+\tilde \rho_{2})/\sqrt{2}\big]=m_{\rm ph}\big[( A_{1}+A_{2})/\sqrt{2}\,\big]=2\Lambda\,,
\eeq
while for the orthogonal combinations we get
\beq
m\big[(\tilde \rho_{1}-\tilde \rho_{2})/\sqrt{2}\big]=m_{\rm ph}\big[( A_{1}-A_{2})/\sqrt{2}\,\big]=2\Lambda\sqrt{1+u}\,.
\label{47}
\eeq

The fermionic part of the heterotic modification in Eq.\,(\ref{Leff}) in terms of the canonical $\tilde\lambda$ fields introduced in 
Eq.\,(\ref{tildelambda}) reduces to
\beq
2\sqrt{u} \,\Lambda\, e^{-i\langle \alpha_{1}\rangle/2}\, i \,\zeta_{R} \,\frac{\tilde\lambda^{\dagger}_{1L}+i\tilde\lambda^{\dagger}_{2L}}{\sqrt{2}} +{\rm H.c.}\,.
\eeq
It implies that the mass of the orthogonal combination $(\tilde\lambda_{1}+i\tilde\lambda_{2})/\sqrt{2}$
is not modified,
\beq
m\big[(\tilde \lambda_{1}+i\tilde \lambda_{2})/\sqrt{2}\big]=2\Lambda\,,
\eeq
The field $\tilde\lambda_{-R}=(\tilde\lambda_{1R}-i\tilde\lambda_{2R})/\sqrt{2}$ mixes with $\zeta_{R}$,
forming 
\beq
\frac{\tilde\lambda_{-R}+\sqrt{u} \,e^{-i\langle \alpha_{1}\rangle/2} \,\zeta_{R}}{\sqrt{1+u}}
\label{l+}
\eeq
under diagonalization. In conjunction with the field $\tilde\lambda_{-L}$ it results in the mass 
\beq
m\Big[\tilde\lambda_{-L},\big(\tilde\lambda_{-R}+\sqrt{u} \,e^{i\langle \alpha_{1}\rangle/2} \,\zeta_{R}\big)/\sqrt{1+u}\Big]=2\Lambda\sqrt{1+u}\,,
\eeq
i.e. the same as in (\ref{47}).

The combination 
\beq
\frac{\zeta_{R} -\sqrt{u}\, e^{i\langle \alpha_{1}\rangle/2} \,\tilde\lambda_{+R}}{\sqrt{1+u}}
\eeq
orthogonal to (\ref{l+}) represents a massless right-moving fermion. It is not a Goldstino fermion, however, since
its residue to the supercurrent vanishes, together with the finishing of $\sum \sigma_{F}$.

Thus, our large-$N$ study of  the connected model mass spectrum demonstrates the following phenomenon.
In addition to the extra massless fermion, we obtain two supermultiplets 
of ${\cal N}=(2,2)$ supersymmetry. The breaking of (2,\,2) supersymmetry in the mass spectrum down to ${\cal N}=(0,2)$ 
does not show up in 
the leading-$N$ approximation at large $N$. The reason for this is  visible in the above derivation: 
the effect of the heterotic modification, say, for fermions, appears just as an admixture of $\zeta$ to $\lambda$,
which does not break the (2,\,2) supersymmetry. 
This feature is not maintained for terms that are cubic, quartic, etc. in fields in the effective action,
implying that breaking of (2,\,2) down to (0,\,2) supersymmetry shows up in  the next order in $1/N$.

The breaking to (0,\,2) in the  $1/N$ corrections also shows up in the running of the $Z$ factors in
the model at hand. This running will be discussed in Sec.\,\ref{befu}.

\section{Pattern of quiver Yang-Mills: constructing  \\  a variety of
 connected sigma models}
\label{otq}

The simplest (0,\,2)  model presented in Sec.\,\ref{htm} can be extended in many distinct ways similar to the pattern
used in four-dimensional Yang-Mills (Fig.\,\ref{f1}).
For instance, the target space (\ref{ts1}) can be expanded up to
\beq
\big[{\rm CP}(N\!-\!1)\big]^n\times \big[{\rm C}^1\big]^n
\label{ts2}
\eeq 
by replacing a single ${\mathcal B}$ superfield by an ensemble of $n$ superfields 
$${\mathcal B}_{12}, \, {\mathcal B}_{23}, \, \ldots {\mathcal B}_{n-1,n},\,{\mathcal B}_{n,1}\,,$$
see Fig.\,\ref{f1}.
The Lagrangian  in the geometric formulation takes the form
\beqn
\mathcal{L}_n
\!&=&\!\!  \sum_{F=1}^n\left\{ -\frac{1}{4} \int \!  d \theta \left[ G_{i\bar j}(A_{F}, A_{F}^{\dagger}) ({\bar{D}A_{F}^{\dagger \bar j})
\,
i\partial_{R}} A_{F}^{i} +{\rm H.c.}\right]\right.
\nonumber\\
&+&\! \! 
\left.
\frac{1}{2}\!\int\! d^2 \theta\,
 G_{i\bar j}(A_{F}, A_{F}^{\dagger})\, B_{F}^{\dagger\bar j} B_{F}^{i } 
 \rule{0mm}{5mm} \right\} +\frac{1}{2}\sum_{F=1}^n \int d^2 \theta \,{\mathcal B}^{\dagger}_{F,F+1}
 {\mathcal B}_{F,F+1}
\label{31}\\
&-&\! \! 
\frac{\kappa}{2}\sum_{F=1}^n \int \!  d \theta\left\{ G_{i\bar j}(A_{F}, A_{F}^{\dagger})  
(\bar{D}A_{F}^{\dagger \bar j})B_{F}^{i}\left({\mathcal B}_{F-1,F}+ {\mathcal B}_{F,F+1}\right) +{\rm H.c.}\right\}\,.
\nonumber
\eeqn

In the gauged formulation we have
\beqn
{\mathcal L}_{n\,\,{\rm gauged}} 
&=&\!\! \sum_F\left\{
 \left| {\mathcal D}_{F\mu} n^i_F\right|^2 - 2|\sigma_F|^2\, |n^i_F|^2
- D_F \left(|n^i_F|^2 -2/g^{2}
\right)\right.
\nonumber\\
&+&\!\!
\big(\xi_{F}^{\dagger}\big)_{jR}\,i{\mathcal D}_{FL}\left(\xi_F\right)^j_R +\big(\xi^{\dagger}_{F}\big)_{jL}\,i{\mathcal D}_{FR}\left(\xi_F\right)^j_L
\nonumber\\[1mm]
&+&\!\!
\left.
\left[\sqrt{2} \sigma_F\big(\xi_{F}^{\dagger}\big)_{jR}\left(\xi_F\right)^j_L +
\sqrt{2}\, n_{Fj}^{\dagger}\big(\lambda_{FR}\xi_{FL}^j +\lambda_{FL}\xi_{FR}^j\big)+{\rm H.c.}\right]
\right\}
\nonumber\\[2mm]
 &+&\!\!\big(\zeta^{\dagger}_{F,F+1}\big)_{\!R} \, i
\partial_L
\, \zeta_R^{F,F+1}\,  - \sum_F\left[
\rule{0mm}{5mm}
 \frac{4\kappa}{g^{2}}\,i\,\lambda_{FL}^{\dagger}\!\left(\zeta_R^{F-1,F} +\zeta_R^{F,F+1}\right)
  + {\rm H.c.}\right] 
 \nonumber \\
 &-& \! \!
 \frac{8|\kappa|^{2}}{g^{4}}\,\sum_{F=1}^n \left| \sigma_{F-1}+\sigma_F\right|^2\,.
 \label{32}
\eeqn

For even values of $n$, this model also has supersymmetric vacua with vanishing energy, i.e. 
${\mathcal N}=(0,2)$ is unbroken. To this end --- keeping supersymmetry unbroken ---
one should choose the set of the vacuum values of $\sigma_F$ to be sign alternating, e.g.,
$(1/2)\left(\Lambda^2,\,-\Lambda^2,\, \Lambda^2,\,-\Lambda^2,...\right)$.
The large\,-$N$ solution can be obtained along the same lines as in Sec.\,\ref{htm}.

\section{Witten's index and its generalization}
\label{WI}

An investigation of the Witten index in a general class of (0,\,2) models was carried out in \cite{BEHT}. In our case
the Witten index vanishes for all connected sigma models but permutation symmetries of the models
allow us to introduce a nonvanishing modified index.

The vanishing of the Witten index in  the heterotically modified ${\rm CP}(N\!-\!1)$ was clearly demonstrated  
by the large-$N$ solution  \cite{L1},
where spontaneous breaking of (0,\,2) supersymmetry is explicit.
The connected extensions considered here preserve the feature of the vanishing Witten index.

To see that this is indeed the case,
 let us consider the model on a finite-size circle; i.e., let us compactify the 
 spatial dimension by imposing periodic boundary conditions both on bosons and fermions, 
 which preserves supersymmetry. In the limit when all heterotic couplings $\kappa_{F}$ in Eq.\,(\ref{sfLa}) 
 are small, we have 
the same bosonic vacua as in the unmodified (2,\,2) models. 
For example, in the case of $\prod {\rm CP}(N_{i}\!-\!1)$, the number of these bosonic vacua 
is $\prod N_{i}$. In addition, in the limit of $\kappa_{F}\to 0$, we have a free 
massless fermion field $\zeta_{R}$, which at the finite-size circle
has two zero modes, one for $\zeta_{R}$, another for $\zeta_{R}^{\dagger}$. 
Fermionic operators of creation and annihilation can be introduced in the standard way. 
 The corresponding zero-energy fermion state can be either empty (bosonic vacuum) 
 or once filled (fermionic counterpartner). Therefore, each bosonic and 
 fermionic vacua always come together in the theories of Secs.\,\ref{htm} and \ref{otq}.
 
 The vanishing of the Witten index usually implies that in some higher approximation (e.g., nonperturbatively),
 supersymmetry will be spontaneously broken since there is no apparent robust protection against this breaking.
Such a protection can exist though if there exists a nonvanishing extended flavor index, in the same vein as in \cite{Witten:2005px}. 
In our model, an extra flavor symmetry is the permutation symmetry of the ${\rm CP}(N\!-\!1)$ factors  from
(\ref{thfour}).
 
 For simplicity let us consider the same case as in Sec.\,\ref{vst}, i.e., $n\!=\!2$ and $N$ even. 
 The generalizations are straightforward. 
 Let us choose $\theta_1=0$ and $\theta_2 = 2\pi\,\frac N2$ for the two ${\rm CP}(N\!-\!1)$ factors. 
 Classically we have a discrete symmetry
 \beq 
 \zeta_R \to - \zeta_R\,,\quad \psi_{R}^{f} \to -  \psi_{R}^{f}\,.
 \label{refl}
 \eeq
 At the quantum level, this symmetry is broken, which is visible from the  existence of fermion condensates
 $\langle \psi_{Lf}^{\dagger}\psi_{R}^{f}\rangle$\,.
 However, applying  in addition the permutation of two ${\rm CP}(N\!-\!1)$ factors
 \beq
 \Phi^{1} \leftrightarrow \Phi^{2}
 \label{perm}
 \eeq
 we get an invariance of the theory.
 The combination of (\ref{refl}) and (\ref{perm}) is a good symmetry, which we will call $P$  conjugation.
 Now we introduce a modification of the Witten index $ I_{W}={\rm Tr} (-1)^{F}$ of the form
 \beq
 I_{P}={\rm Tr} \big[P(-1)^{F}\big]
 \label{IP}
 \eeq
 It is clear that the index does not vanish in contradistinction with $I_{W}$: an addition of the $\zeta_{R}$ fermion 
 to the state now yields the positive sign because of $P$.
 
 Another way to show the absence of the spontaneous supersymmetry breaking in the case is as follows.
 The order parameter for supersymmetry breaking is 
 \beq
 \mathcal{F}_\zeta = \mbox {const} \cdot\kappa \left(\sum_{f=1,2} \psi_{L\,f}^+\,\psi_R^f \, G^f
 \right)
 \eeq
 where $G^f$ is the metric of the corresponding ${\rm CP}(N\!-\!1)$ factor.
 
 Now, let us apply the transformation (\ref{refl}). As was mentioned above, this classical symmetry is broken 
 at the quantum level due to the chiral anomaly; it 
 changes the vacuum angles, namely,
 \beq
 \theta_1 \to  2\pi\,\frac N2\,,\qquad  \theta_2 \to  2\pi\,\frac N2+  2\pi\,\frac N2= 2\pi N\,\,\,\mbox{equiv} \,\,0\,.
 \eeq
 Under the rotation above the order parameter
 \beq
  \mathcal{F}_\zeta  \to \exp (i\pi)  \mathcal{F}_\zeta\,.
 \eeq
At the same time, interchanging the two ${\rm CP}(N\!-\!1)$ factors we see that $ \mathcal{F}_\zeta$
remains intact. Combining these two facts, we conclude that $ \mathcal{F}_\zeta=0$.
Note that $ \mathcal{F}_\zeta$ presents also the coupling
of the would-be Goldstino to the supercurrent.  

Another very simple example is $n=N$. In this case, we must choose
\beq
\theta_1=0,\,\, \theta_2= 2\pi,\,\,\theta_3=4\pi,\, ... \,, \theta_N=2\pi (N\!-\!1)\,,
\eeq
 and the rotation 
 \beq
 \psi_R^f\to e^{i2\pi/N}  \psi_R^f \,\, \mbox{ for all }\,\, f\,, \quad \zeta_R \to \zeta_R\exp (-i2\pi/N)\,.
 \eeq
 The same argument as above implies $ \mathcal{F}_\zeta =0$. 
 
\section{Beta functions}
\label{befu}

The $\beta$ functions of the basic heterotic model discussed
in Sec.\,\ref{tshn} were derived in \cite{C2,CCSV}.
In the heterotic model one deals with two coupling constants: $g^2$ appearing in the metric and the deformation parameter $\kappa$. 
The $\beta$ function for $g^2$ is \cite{CCSV}
\beqn
\beta_{g}\!=\!\mu\,\frac{dg^{2}}{d\mu}\!=\!-\frac{g^{2}}{4\pi}\,\frac{T_{G}\,g^{2}\left(1+\gamma_{\psi_R}/2
\right) - {h}^{2}\left(\gamma_{\psi_{R}}+\gamma_{\zeta}\right)}{1-({h}^{2}/4\pi)}\,,
\label{11}
\label{betagh}
\eeqn
where $\gamma_{\zeta}=-\mu \,d\log {\cal Z}/d\mu$ and $\gamma_{\psi_R}=-\mu \,d\log  Z/d\mu$ are the anomalous dimensions of the
corresponding fields, which to the leading order are proportional to the coupling
\beq
h^{2}=\frac{|\kappa|^{2}}{Z{\mathcal Z}}\,.
\label{ai15}
\eeq
Here, $Z$ and ${\mathcal Z}$ are field renormalization constants for $\psi_R$ and $\zeta_R$, respectively 
[see Eqs.\,(\ref{sfLa}) and (\ref{components}) for their definition]. At one  loop \cite{C2},
\beq
 \gamma^{(1)}_{\psi_{R}}=\frac{{h}^{2}}{2\pi}\,,\qquad \gamma^{(1)}_{\zeta}=\frac{(N-1)\,{h}^{2}}{2\pi}\,.
 \label{ai16}
\eeq

Now, in the connected models, the general relation (\ref{betagh}) remains intact, while the 
expression for the anomalous dimension $\gamma$ changes. In particular, for the model of Sec.\,\ref{htm}
\beq
\gamma^{(1)}_{\psi_{R}}=\frac{{h}^{2}}{2\pi}\,,\qquad  \gamma^{(1)}_{\zeta} = n\frac{(N-1)\,{h}^{2}}{2\pi}\,,
\eeq
and $T_{G}=N$.

The $\beta$ function for $h^{2}$ is also fixed by anomalous dimensions; see \cite{CCSV,C2} for details.
There is a fixed point for the ratio $h^{2}/g^{2}$,
\beq
\frac{h^{2}}{g^{2}}\Bigg|_{c}=\frac 1 2\cdot\frac{N}{n\,(N-1)+1}\,.
\eeq
At large $n$ it means that $nh^{2}$ and $g^{2}$ scale exactly in the same way (as it occurred at one loop).

 \section{Conclusions}
 \label{concl}

 In this paper we suggested a way to make connected two-dimensional ${\mathcal N}\! =\!(0,2)$ 
 sigma models from ${\mathcal N}\!=\!(2,2)$ CP$(N\!-\!1)$ models. This method is easily extendible 
 to any Grassmannian sigma model. To this end, one introduces an extra fermion 
 \mbox{${\mathcal N}\! =\!(0,2)$} superfield (or superfields)  on $C^1$ coupled to all or some of 
 the $n$ copies of the \mbox{${\mathcal N}\!=\!(2,2)$} sigma model. The connected model emerging 
 in this way can be solved in the large-$N$ limit. Our solution demonstrated that ${\mathcal N} \!=\!(0,2)$  
 supersymmetry which is spontaneously broken without
``connection," is restored in the connected version.  
This statement is unambiguously proved in the leading order in $1/N$.

Then, in Sec.\,\ref{WI}, we introduced a generalized Witten index which, being nonvanishing, provided us with the 
general proof of the
exact statement: our connected models do have supersymmetric vacua. 

We also  found the excitation spectrum in the leading $1/N$ approximation and expressions for the beta 
functions in the quiver models.
 
 \vspace{1cm}
 
 \section*{Acknowledgments}
We thank S. Gukov and P. Putrov for helpful discussions.

The work of M.\,S. is supported in part by DOE Grant DE-SC0011842.
Kind hospitality and support extended to M.\,S. during his stay at Institut des Hautes \'Etudes Scientifiques is acknowledged.
A.\,V. appreciates hospitality of the Kavli Institute for Theoretical Physics where his research was supported in part by the National Science Foundation under Grant No.\ NSF PHY11-25915. 
His work was also supported in part by the National Science Foundation under Grant No. PHYS-1066293 and the hospitality 
of the Aspen Center for Physics.
The work of A.\,Y. was  supported by  William I. Fine Theoretical Physics Institute, University of Minnesota, 
by RFBR Grant No.\ 13-02-00042a, and by Russian State Grant for 
Scientific Schools RSGSS-657512010.2. The work of A.\,Y. was also supported 
by RSCF Grant No. 14-22-0021.

\end{document}